\pdfoutput=1
\documentclass[preprint,preprintnumbers,amsmath,amssymb,floatfix,endfloats*]{revtex4}

\usepackage{graphicx}
\usepackage{dcolumn}
\usepackage{bm}
\usepackage{amsfonts}

\DeclareMathAlphabet{\mathsfsl}{OT1}{cmr}{bx}{it}
\begin{document}
\title{The effect of cryogenic thermal cycling on aging, rejuvenation,
and mechanical properties of metallic glasses}
\author{Nikolai V. Priezjev$^{1,2}$}
\affiliation{$^{1}$Department of Mechanical and Materials
Engineering, Wright State University, Dayton, OH 45435}
\affiliation{$^{2}$National Research University Higher School of
Economics, Moscow 101000, Russia}
\date{\today}
\begin{abstract}

The structural relaxation, potential energy states, and mechanical
properties of a model glass subjected to thermal cycling are
investigated using molecular dynamics simulations. We study a
non-additive binary mixture which is annealed with different cooling
rates from the liquid phase to a low temperature well below the
glass transition. The thermal treatment is applied by repeatedly
heating and cooling the system at constant pressure, thus
temporarily inducing internal stresses upon thermal expansion. We
find that poorly annealed glasses are relocated to progressively
lower levels of potential energy over consecutive cycles, whereas
well annealed glasses can be rejuvenated at sufficiently large
amplitudes of thermal cycling.  Moreover, the lowest levels of
potential energy after one hundred cycles are detected at a certain
temperature amplitude for all cooling rates.   The structural
transition to different energy states is accompanied by collective
nonaffine displacements of atoms that are organized into clusters,
whose typical size becomes larger with increasing cooling rate or
temperature amplitude.   We show that the elastic modulus and the
peak value of the stress overshoot exhibit distinct maxima at the
cycling amplitude, which corresponds to the minimum of the potential
energy. The simulation results indicate that the yielding peak as a
function of the cycling amplitude for quickly annealed glasses
represents a lower bound for the maximum stress in glasses prepared
with lower cooling rates.

\vskip 0.5in


\end{abstract}

\maketitle

\section{Introduction}

The development of novel processing routes that include thermal and
mechanical treatments of metallic glasses is important for numerous
engineering applications~\cite{Kruzic16}. It is well accepted by now
that in contrast to crystalline materials, amorphous solids can
deform plastically via the so-called shear transformations, which
involve only a few tens of atoms~\cite{Argon79,Spaepen77}.
Remarkably, recent experimental studies on metallic glasses have
demonstrated that cryogenic thermal cycling can induce rejuvenation
that leads to less relaxed states of higher energy and improved
plasticity~\cite{Ketov15,Lu18,Kerscher18,Saida18,Ketov18}.  It was
argued that atomic-scale structural rejuvenation upon thermal
cycling might be caused by internal stresses that arise due to
spatially heterogeneous thermal expansion of the amorphous
material~\cite{Hufnagel15}.  More recently, the probability
distribution of local thermal expansion coefficients was estimated
using atomistic simulations, and it was shown that internal stress
can exceed the local yield stress and therefore trigger local shear
transformations~\cite{Barrat18}.   Similar to thermal agitation, it
was originally demonstrated that a shear cycle can either overage or
rejuvenate the glass depending on whether the strain amplitude is
smaller or larger than the yield strain~\cite{Lacks04}.  However,
despite extensive efforts, the underlying mechanisms of structural
relaxation or rejuvenation during thermomechanical treatments are
still not fully understood.

\vskip 0.05in

In the last few years, the evolution of the potential energy,
structural relaxation dynamics,  and mechanical properties of
glasses subjected to multiple shear cycles were examined using
molecular dynamics (MD)
simulations~\cite{Priezjev13,Sastry13,Reichhardt13,Priezjev14,IdoNature15,Priezjev16,Kawasaki16,
Priezjev16a,Sastry17,Priezjev17,OHern17,Hecke17,Priezjev18,Priezjev18a,Sastry18}.
In particular, it was found that progressively lower energy states
can be reached even for poorly annealed glasses by imposing a
periodic shear deformation with the strain amplitude below
yield~\cite{Sastry17,Priezjev18,Priezjev18a,Sastry18}. As the cycle
number increases, the gradual decay of the potential energy at
finite temperature is reflected in a reduction in size of transient
clusters of atoms with relatively large nonaffine
displacements~\cite{Priezjev18,Priezjev18a}. Upon approaching the
yielding point from below, the potential energy decreases, while the
storage modulus is slightly reduced~\cite{Priezjev18a}.  Above the
yielding transition, periodically sheared glasses settle at higher
levels of potential energy, the loss modulus increases, and a shear
band is formed after a number of cycles in sufficiently large
systems~\cite{Sastry17,Priezjev17,Priezjev18a}.   However, the
optimal parameter values for the strain amplitude, oscillation
period, temperature, and cooling rate required to access a wide
range of energy states and tune mechanical properties still need to
be determined.

\vskip 0.05in

In this paper, the influence of thermal cycling on the potential
energy and mechanical properties of binary glasses is studied using
molecular dynamics simulations. The glasses are initially annealed
across the glass transition with various cooling rates to a low
temperature and then subjected to repeated heating and cooling
cycles at constant pressure. We show that quickly annealed glasses
reach lower levels of potential energy after one hundred thermal
cycles, while slowly annealed glasses are rejuvenated at large
temperature amplitudes.  The results of numerical simulations
indicate that the elastic modulus and the yield stress as a function
of the cycling amplitude each exhibit a distinct maximum, which
correlates well with the lowest potential energy level detected at
the same amplitude.

\vskip 0.05in

This paper is organized as follows. The details of molecular
dynamics simulations including parameter values, equilibration
procedure, and temperature protocol are described in the next
section.   The results for the potential energy series, analysis of
nonaffine displacements, and stress-strain response are presented in
section\,\ref{sec:Results}. A brief summary is provided in the last
section.

\section{Details of MD simulations}
\label{sec:MD_Model}

\begin{figure}[t]
\includegraphics[width=10.0cm,angle=0]{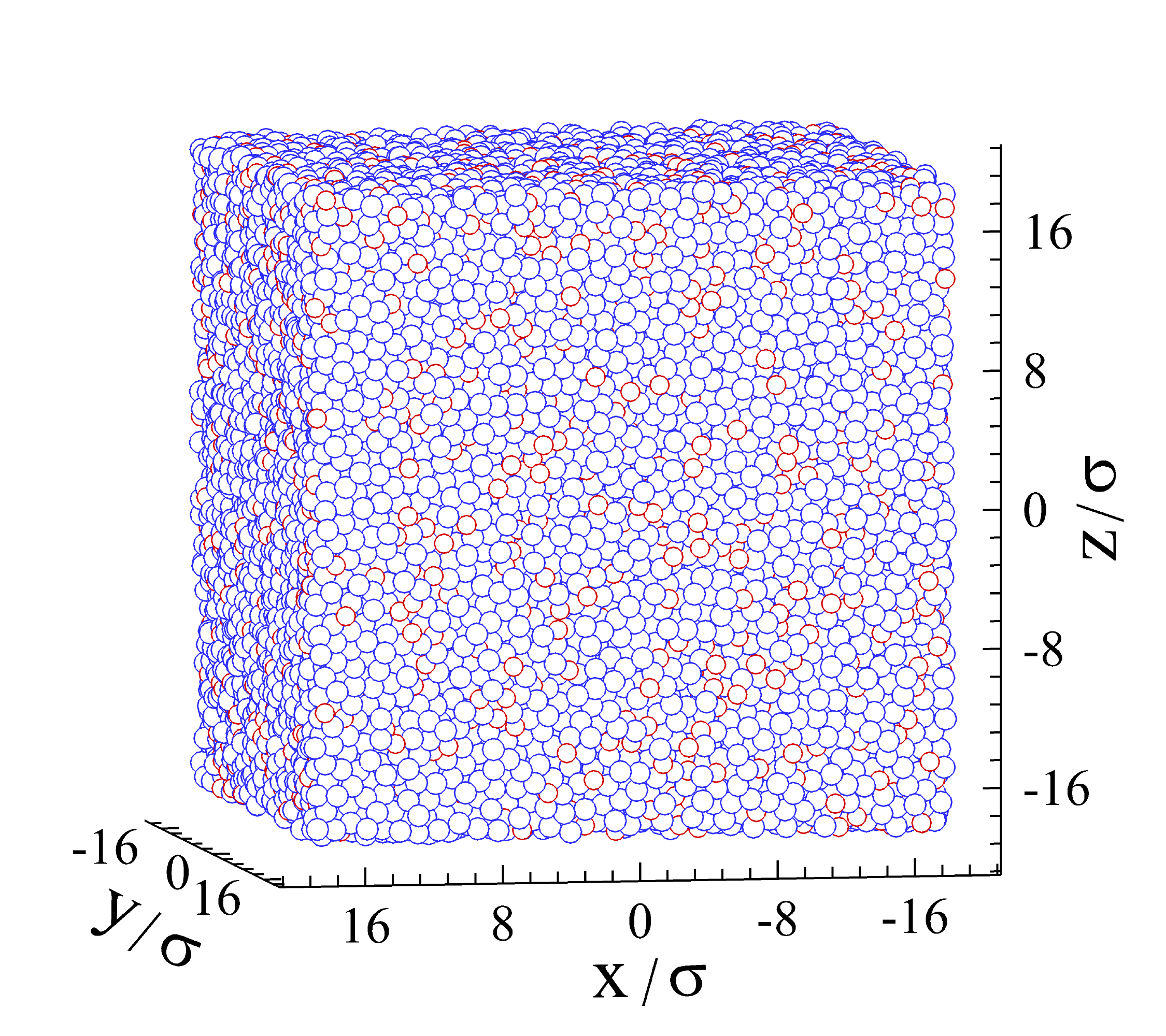}
\caption{(Color online) An atomic configuration of the binary
Lennard-Jones glass ($N_{tot}=60\,000$) after isobaric quench from a
liquid state to the temperature $T_{LJ}=0.01\,\varepsilon/k_B$ with
the cooling rate of $10^{-3}\varepsilon/k_{B}\tau$. Atoms of types
$A$ and $B$, denoted by blue and red spheres, are not shown to
scale.}
\label{fig:snapshot_system}
\end{figure}

The model glass studied in this paper is represented by the binary
Lennard-Jones (LJ) mixture (80:20), which was originally developed
by Kob and Andersen (KA)~\cite{KobAnd95} to study the amorphous
metal alloy $\text{Ni}_{80}\text{P}_{20}$~\cite{Weber85}. In this
model, any two atoms of types $\alpha,\beta=A,B$ interact with each
other via the truncated LJ potential:
\begin{equation}
V_{\alpha\beta}(r)=4\,\varepsilon_{\alpha\beta}\,\Big[\Big(\frac{\sigma_{\alpha\beta}}{r}\Big)^{12}\!-
\Big(\frac{\sigma_{\alpha\beta}}{r}\Big)^{6}\,\Big],
\label{Eq:LJ_KA}
\end{equation}
where the interaction parameters for both types of atoms are
specified as $\varepsilon_{AA}=1.0$, $\varepsilon_{AB}=1.5$,
$\varepsilon_{BB}=0.5$, $\sigma_{AB}=0.8$, $\sigma_{BB}=0.88$, and
$m_{A}=m_{B}$~\cite{KobAnd95}.  The LJ potential is truncated at the
cutoff radius $r_{c,\,\alpha\beta}=2.5\,\sigma_{\alpha\beta}$ to
reduce computational cost.   The reduced LJ units of length, mass,
energy, and time were used to express physical quantities; more
specifically, $\sigma=\sigma_{AA}$, $m=m_{A}$,
$\varepsilon=\varepsilon_{AA}$, and
$\tau=\sigma\sqrt{m/\varepsilon}$.   The total number of atoms in
the system is $N_{tot}=60\,000$.   A representative snapshot of the
atomic configuration in the glassy phase is shown in
Fig.\,\ref{fig:snapshot_system}.   The equations of motion for each
atom were solved numerically using the velocity Verlet
algorithm~\cite{Allen87} with the time step $\triangle
t_{MD}=0.005\,\tau$ using the open-source LAMMPS code~\cite{Lammps}.

\vskip 0.05in


In our setup, all atoms were initially placed in a periodic box and
the system was allowed to equilibrate at a high-temperature liquid
state in the $NPT$ ensemble. Throughout the study, the pressure was
kept constant, \textit{i.e.}, $P=0$.   The temperature, denoted by
$T_{LJ}$, was regulated using the Nos\'{e}-Hoover
thermostat~\cite{Lammps}. The computer glass transition temperature
for the KA model is $T_c\approx0.435\,\varepsilon/k_B$ at the atomic
density $\rho=\rho_{A}+\rho_{B}=1.2\,\sigma^{-3}$~\cite{KobAnd95}.
Note that $k_B$ denotes the Boltzmann constant.  In the next step,
the system was thermally annealed across the glass transition to the
target temperature of $0.01\,\varepsilon/k_B$ with cooling rates
$10^{-2}\varepsilon/k_{B}\tau$, $10^{-3}\varepsilon/k_{B}\tau$,
$10^{-4}\varepsilon/k_{B}\tau$, and $10^{-5}\varepsilon/k_{B}\tau$.
Thus, the typical annealing time varies from about $100\,\tau$,
which is only a couple of orders of magnitude greater than the
atomic vibration time, to about $10^5\,\tau$, which requires an
efficient parallel code for the large system size~\cite{Lammps}.
Next, the annealed glasses were thermally cycled with the period
$T=10\,000\,\tau$ over 100 cycles.  During production runs, the
system dimensions, temperature, pressure tensor, potential energy,
as well as atom positions at the end of each cycle were saved for
the post-processing analysis.

\section{Results}
\label{sec:Results}

\begin{figure}[t]
\includegraphics[width=12.0cm,angle=0]{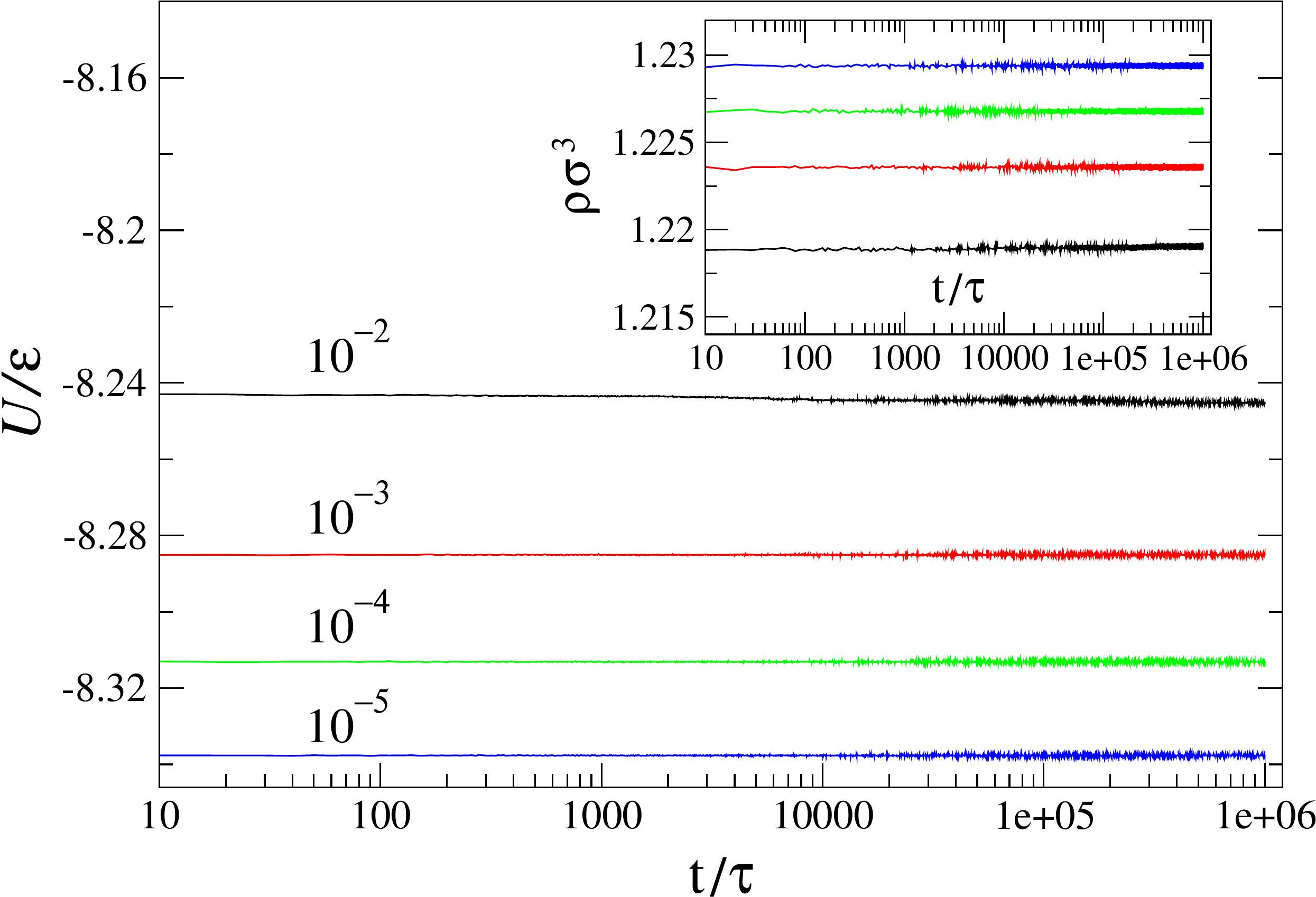}
\caption{(Color online) The potential energy during aging process at
$T_{LJ}=0.01\,\varepsilon/k_B$ in samples prepared with cooling
rates $10^{-2}\varepsilon/k_{B}\tau$ (black curve),
$10^{-3}\varepsilon/k_{B}\tau$ (red curve),
$10^{-4}\varepsilon/k_{B}\tau$ (green curve), and
$10^{-5}\varepsilon/k_{B}\tau$ (blue curve). The inset shows the
evolution of the glass density for the same samples. }
\label{fig:poten_dens_T0.01}
\end{figure}

We first discuss the variation of the potential energy and glass
density for samples annealed with cooling rates
$10^{-2}\varepsilon/k_{B}\tau$, $10^{-3}\varepsilon/k_{B}\tau$,
$10^{-4}\varepsilon/k_{B}\tau$, and $10^{-5}\varepsilon/k_{B}\tau$
to the low temperature of $0.01\,\varepsilon/k_B$ and aged at this
temperature and zero pressure for the time interval of $10^6\tau$.
The results are summarized in Fig.\,\ref{fig:poten_dens_T0.01}.   As
expected, more slowly annealed glasses at constant pressure attain
lower levels of potential energy and become denser. Moreover, there
is a noticeable decay in the potential energy for the poorly
annealed glass (cooling rate $10^{-2}\varepsilon/k_{B}\tau$),
whereas the potential energy for more slowly cooled glasses remains
essentially constant during the time interval $10^6\tau$. In what
follows, these curves will be used as references for the analysis of
potential energy series in thermally cycled glasses. Furthermore, as
shown in the inset to Fig.\,\ref{fig:poten_dens_T0.01}, the glass
density for the sample prepared with the fastest cooling rate of
$10^{-2}\varepsilon/k_{B}\tau$ becomes slightly higher after
$10^6\tau$. Apart from thermal fluctuations, the density of the
other samples remains unchanged.   In the previous studies on the KA
model, it was shown that when the system is quenched below the glass
transition at constant volume, the potential energy exhibits a slow
decay over time and the two-times correlation functions show a
universal dependence on the waiting time, indicating progressively
slower particle dynamics~\cite{KobBar97,KobBar00}.

\vskip 0.05in

\begin{figure}[t]
\includegraphics[width=12.0cm,angle=0]{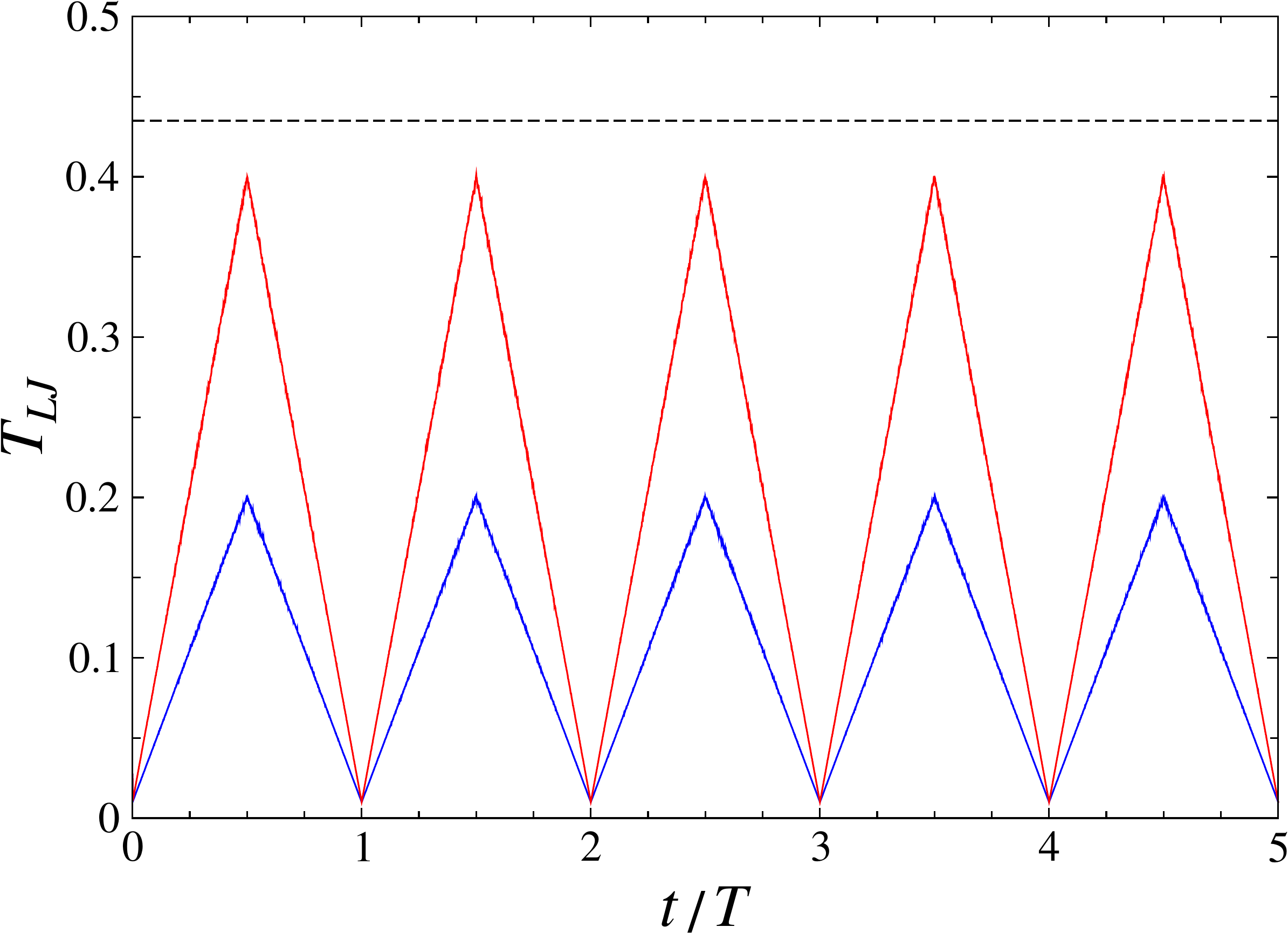}
\caption{(Color online) The variation of temperature $T_{LJ}$ (in
units of $\varepsilon/k_B$) during the first five periods,
$T=10\,000\,\tau$, for the temperature amplitudes
$0.2\,\varepsilon/k_B$ (blue curve) and $0.4\,\varepsilon/k_B$ (red
curve). The data are taken in the sample initially prepared with the
cooling rate of $10^{-4}\varepsilon/k_{B}\tau$. The black dashed
line denotes the critical temperature of $0.435\,\varepsilon/k_B$. }
\label{fig:temp_control}
\end{figure}

After the annealing procedure, the binary glasses were subjected to
repeated thermal cycling, where the temperature was varied piecewise
linearly from $0.01\,\varepsilon/k_B$ to the maximum value and back
to $0.01\,\varepsilon/k_B$ over 100 cycles with the period
$T=10\,000\,\tau$. An example of the temperature profiles measured
during the first five cycles is presented in
Fig.\,\ref{fig:temp_control} for the amplitudes
$0.2\,\varepsilon/k_B$ and $0.4\,\varepsilon/k_B$. It should be
noted that since the simulations are performed at constant pressure
($P=0$), the volume becomes expanded at elevated temperatures (not
shown).   The following temperature amplitudes were considered
$T_{LJ}=0.1\,\varepsilon/k_B$, $0.2\,\varepsilon/k_B$,
$0.3\,\varepsilon/k_B$, $0.4\,\varepsilon/k_B$, and
$0.435\,\varepsilon/k_B$.  Recall that the last value,
$0.435\,\varepsilon/k_B$, is the critical temperature of the KA
model~\cite{KobAnd95}. In this case, the system temporarily enters a
state with sluggish dynamics and then is cooled back to
$0.01\,\varepsilon/k_B$ during each cycle.

\vskip 0.05in

\begin{figure}[t]
\includegraphics[width=12.0cm,angle=0]{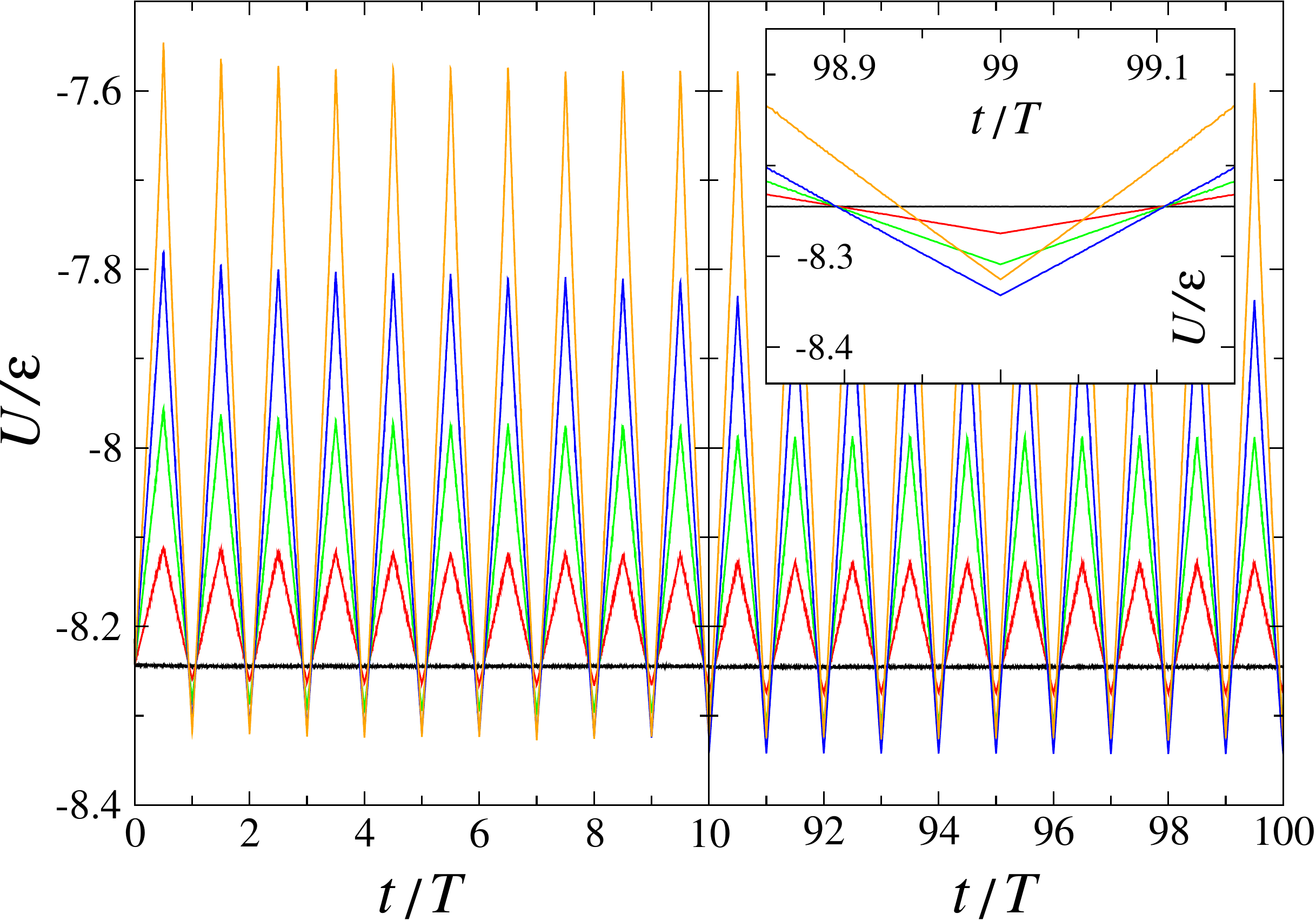}
\caption{(Color online) The potential energy during the first 10
(left panel) and the last 10 (right panel) thermal cycles with
amplitudes $0.1\,\varepsilon/k_B$ (red), $0.2\,\varepsilon/k_B$
(green), $0.3\,\varepsilon/k_B$ (blue), and $0.4\,\varepsilon/k_B$
(orange). The black curve denotes the data for the same sample aged
at the temperature of $0.01\,\varepsilon/k_B$. The cooling rate is
$10^{-2}\varepsilon/k_{B}\tau$. The inset shows an enlarged view of
the same data at $t=99\,T$. }
\label{fig:poten_10m2}
\end{figure}
\begin{figure}[t]
\includegraphics[width=12.0cm,angle=0]{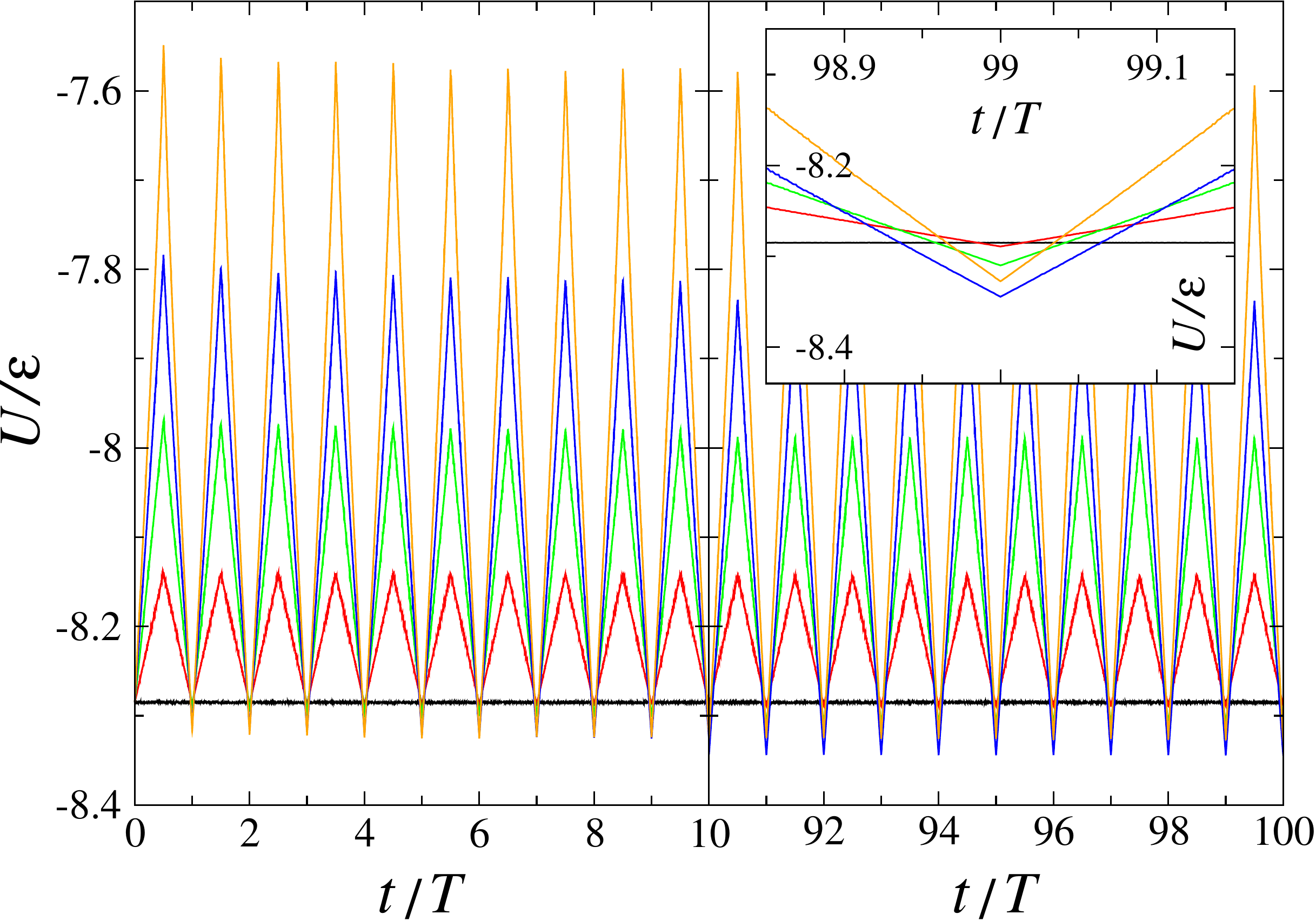}
\caption{(Color online) The potential energy series for the glass
annealed with the cooling rate of $10^{-3}\varepsilon/k_{B}\tau$ and
thermally cycled with amplitudes $0.1\,\varepsilon/k_B$ (red),
$0.2\,\varepsilon/k_B$ (green), $0.3\,\varepsilon/k_B$ (blue), and
$0.4\,\varepsilon/k_B$ (orange). The data at the temperature of
$0.01\,\varepsilon/k_B$ are denoted by the black line. An enlarged
view of the same data is shown in the inset. }
\label{fig:poten_10m3}
\end{figure}
\begin{figure}[t]
\includegraphics[width=12.0cm,angle=0]{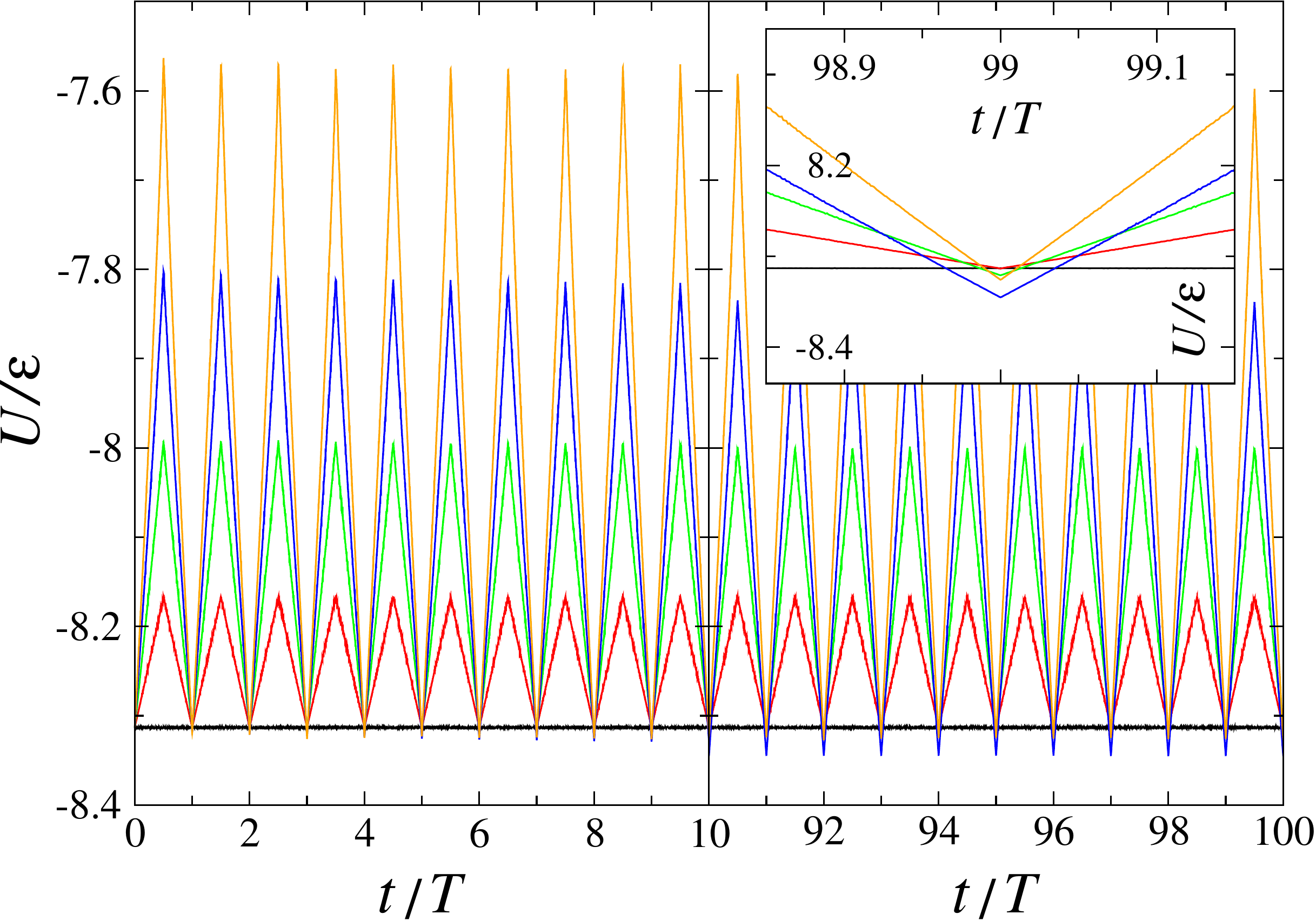}
\caption{(Color online) The variation of the potential energy for
the sample prepared with the cooling rate of
$10^{-4}\varepsilon/k_{B}\tau$ and cycled with thermal amplitudes
$0.1\,\varepsilon/k_B$ (red), $0.2\,\varepsilon/k_B$ (green),
$0.3\,\varepsilon/k_B$ (blue), and $0.4\,\varepsilon/k_B$ (orange).
The black line denotes the potential energy at
$T_{LJ}=0.01\,\varepsilon/k_B$. The inset shows an expanded view of
the same data during the last cycle. }
\label{fig:poten_10m4}
\end{figure}
\begin{figure}[t]
\includegraphics[width=12.0cm,angle=0]{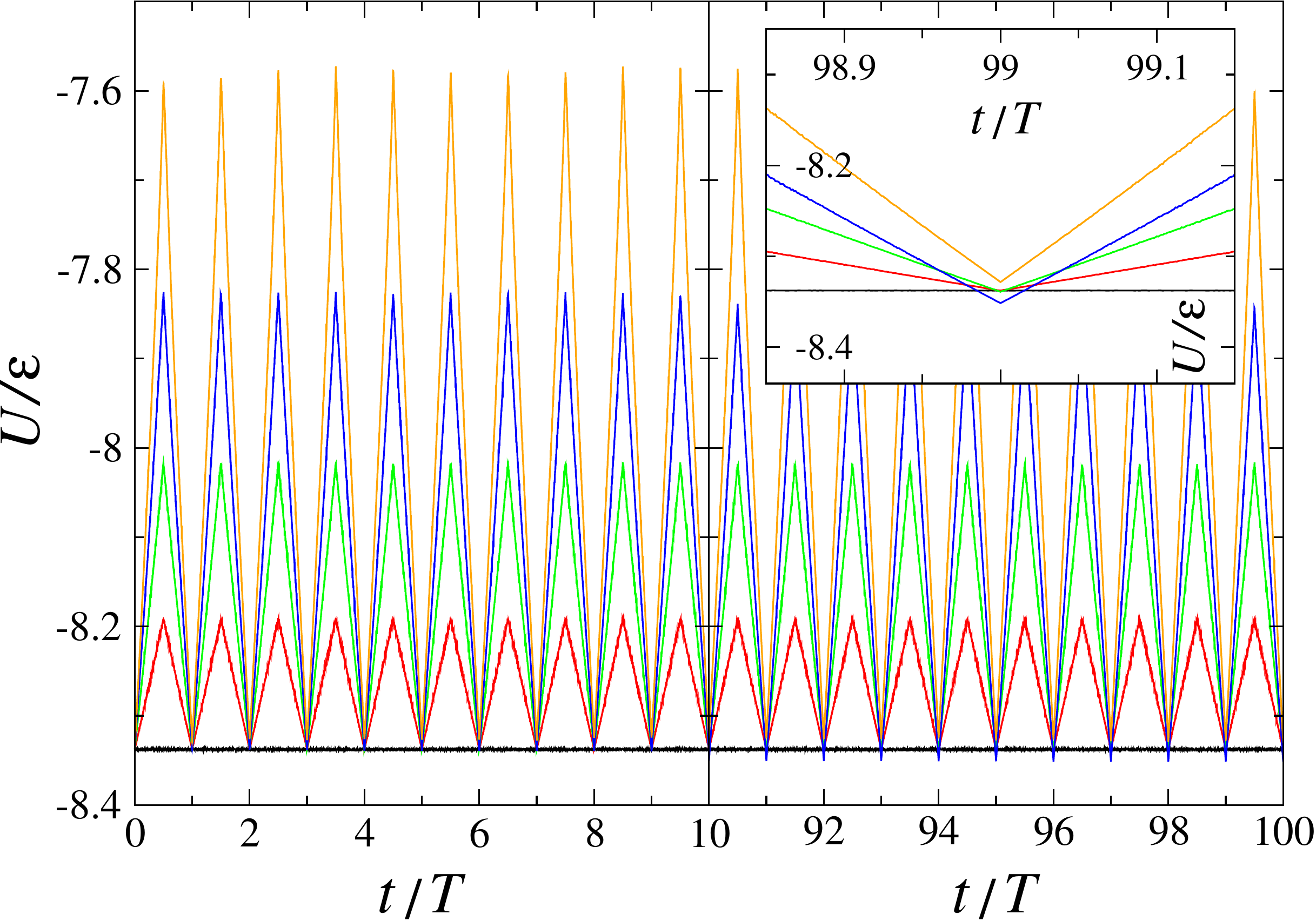}
\caption{(Color online) The potential energy for the well annealed
glass (cooling rate $10^{-5}\varepsilon/k_{B}\tau$) and temperature
amplitudes $0.1\,\varepsilon/k_B$ (red), $0.2\,\varepsilon/k_B$
(green), $0.3\,\varepsilon/k_B$ (blue), and $0.4\,\varepsilon/k_B$
(orange). The potential energy at $T_{LJ}=0.01\,\varepsilon/k_B$ is
indicated by the black line. A close-up view of the same data at
$t=99\,T$ is shown in the inset. }
\label{fig:poten_10m5}
\end{figure}

The potential energy series for thermal cycling with various
amplitudes are presented in Figs.\,\ref{fig:poten_10m2},
\ref{fig:poten_10m3}, \ref{fig:poten_10m4}, and \ref{fig:poten_10m5}
for the glasses annealed with cooling rates
$10^{-2}\varepsilon/k_{B}\tau$, $10^{-3}\varepsilon/k_{B}\tau$,
$10^{-4}\varepsilon/k_{B}\tau$, and $10^{-5}\varepsilon/k_{B}\tau$,
respectively.  For clarity, the data are only shown for the first
and last 10 cycles.  Note that the scale is the same in all four
figures.  It can be clearly observed in
Figs.\,\ref{fig:poten_10m2}--\ref{fig:poten_10m5} that with
increasing maximum cycling temperature, the amplitude of the
potential energy variations becomes greater as the systems get
thermally expanded and the interatomic distances increase on
average. More importantly, however, one can notice that the
potential energy minimum at the end of each cycle depends on the
cooling rate, temperature amplitude, and the cycle number. Thus, in
the case of the poorly annealed glass, shown in
Fig.\,\ref{fig:poten_10m2}, the minima of the potential energy
become progressively lower over consecutive cycles and remain below
the energy level that corresponds to the glass aged at
$T_{LJ}=0.01\,\varepsilon/k_B$.  A more subtle feature of the
potential energy variation during the last cycle can be identified
in the inset to Fig.\,\ref{fig:poten_10m2}; namely, the lowest
energy minimum, $U\approx-8.34\,\varepsilon$, is attained at the
cycling amplitude of $0.3\,\varepsilon/k_B$.

\vskip 0.05in


Similar patterns for the potential energy series can be observed for
more slowly annealed glasses (cooling rates
$10^{-3}\varepsilon/k_{B}\tau$ and $10^{-4}\varepsilon/k_{B}\tau$)
shown in Figs.\,\ref{fig:poten_10m3} and \ref{fig:poten_10m4},
except that thermal cycling with the amplitude
$0.1\,\varepsilon/k_B$ results in the energy minima being only
slightly lower than the energy levels of the glasses aged at
$T_{LJ}=0.01\,\varepsilon/k_B$. Furthermore, as shown in
Fig.\,\ref{fig:poten_10m5}, the well annealed glass (cooling rate
$10^{-5}\varepsilon/k_{B}\tau$) attains nearly the same energy
minima when cycled with the amplitudes $0.1\,\varepsilon/k_B$ and
$0.2\,\varepsilon/k_B$, while the lowest energy minimum,
$U\approx-8.35\,\varepsilon$, is reached at $0.3\,\varepsilon/k_B$.
Interestingly, the well annealed sample becomes rejuvenated at the
cycling amplitude of $0.4\,\varepsilon/k_B$, as its potential energy
minimum is slightly higher than the energy level for the glass aged
at $T_{LJ}=0.01\,\varepsilon/k_B$ (see inset to
Fig.\,\ref{fig:poten_10m5}).  Overall, the important conclusion from
Figs.\,\ref{fig:poten_10m2}--\ref{fig:poten_10m5} is that, despite
the energy levels for annealed glasses in
Fig.\,\ref{fig:poten_dens_T0.01} being distinctly different, the
energy minima for glasses cycled with amplitude
$0.3\,\varepsilon/k_B$ over 100 cycles are nearly the same;
$U(100\,T)$ only slightly decreases from
$U\approx-8.34\,\varepsilon$ for the poorly annealed glass to
$U\approx-8.35\,\varepsilon$ for the well annealed glass.


\vskip 0.05in


Structural relaxation processes in thermally cycled glasses can be
studied by analyzing the so-called nonaffine displacements, which
represent a deviation from the local linear deformation of the
material. Numerically, the nonaffine measure associated with an atom
$i$ is computed via the transformation matrix $\mathbf{J}_i$ that
describes a linear transformation of a small group of neighboring
atoms during the time interval $\Delta t$, while at the same time
minimizing the quantity $D^2(t, \Delta t)$ as follows~\cite{Falk98}:
\begin{equation}
D^2(t, \Delta t)=\frac{1}{N_i}\sum_{j=1}^{N_i}\Big\{
\mathbf{r}_{j}(t+\Delta t)-\mathbf{r}_{i}(t+\Delta t)-\mathbf{J}_i
\big[ \mathbf{r}_{j}(t) - \mathbf{r}_{i}(t)    \big] \Big\}^2,
\label{Eq:D2min}
\end{equation}
where the sum is over nearest neighbors that are located closer than
$1.5\,\sigma$ to the $i$-th atom.   Recently, the analysis of
spatial configurations of atoms with large nonaffine displacements
was carried out for cyclically
sheared~\cite{Priezjev16,Priezjev16a,Priezjev17,Priezjev18,Priezjev18a}
and compressed~\cite{NVP18strload} glasses.   In particular, it was
demonstrated that after a certain number of shear cycles, the
yielding transition at finite temperature is accompanied by the
formation of a system-spanning shear band of large nonaffine
displacements in both well~\cite{Priezjev17} and
poorly~\cite{Priezjev18a} annealed glasses. On the other hand, below
the yield point, binary glasses start to deform reversibly after
transient cycles and nonaffine displacements become organized into
compact
clusters~\cite{Priezjev16,Priezjev16a,Priezjev17,Priezjev18,Priezjev18a,NVP18strload}.

\vskip 0.05in

\begin{figure}[t]
\includegraphics[width=12.cm,angle=0]{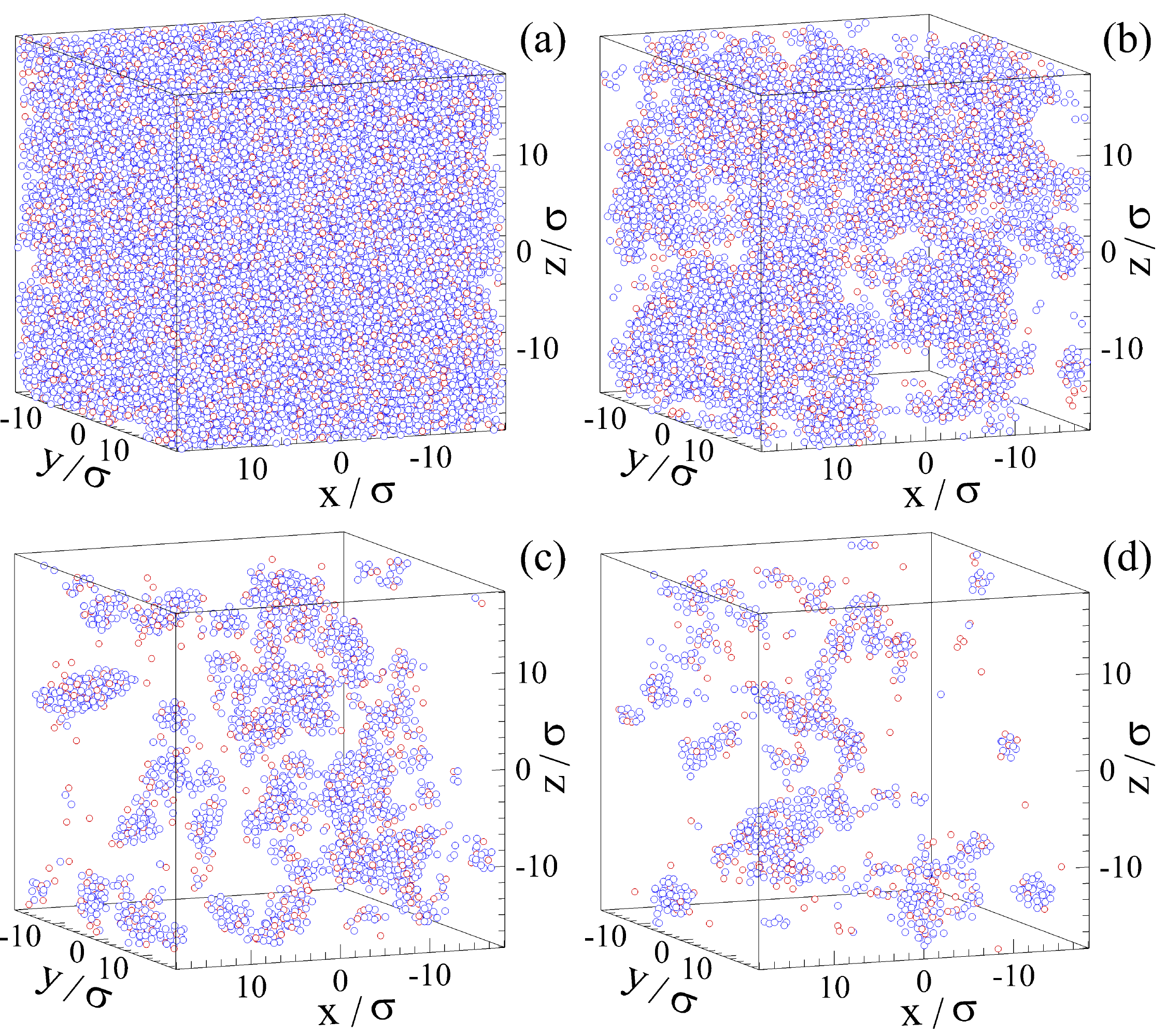}
\caption{(Color online)  Spatial configurations of atoms with large
nonaffine measure (a) $D^2(0,T)>0.04\,\sigma^2$, (b)
$D^2(T,T)>0.04\,\sigma^2$, (c) $D^2(9T,T)>0.04\,\sigma^2$, and (d)
$D^2(99T,T)>0.04\,\sigma^2$. The cooling rate is
$10^{-2}\varepsilon/k_{B}\tau$ and the amplitude of thermal cycling
is $0.2\,\varepsilon/k_B$.  }
\label{fig:snapshot_clusters_Tm02_r10m2}
\end{figure}
\begin{figure}[t]
\includegraphics[width=12.cm,angle=0]{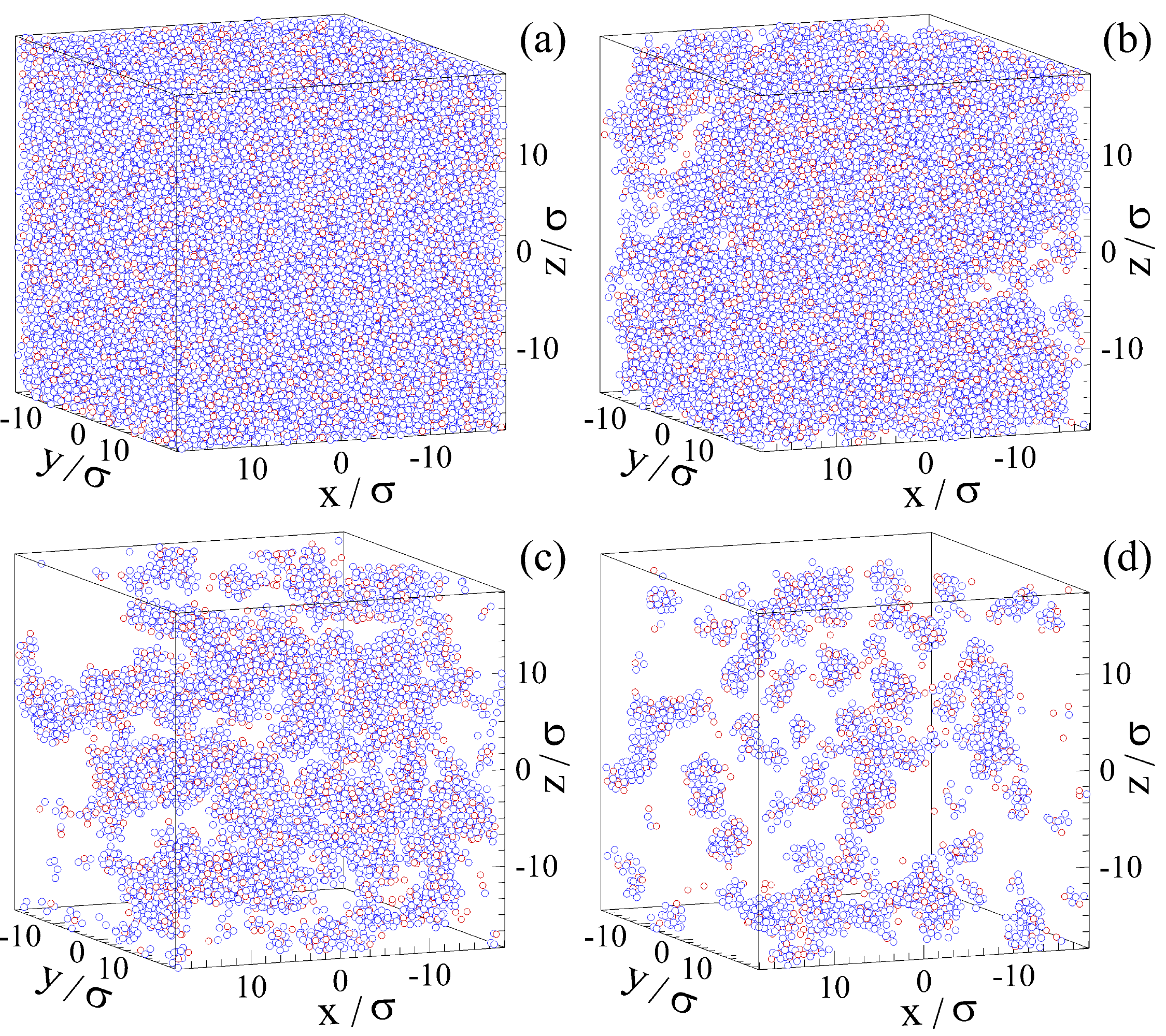}
\caption{(Color online) Atomic configurations of the binary glass
prepared with the cooling rate of $10^{-2}\varepsilon/k_{B}\tau$ and
subjected to thermal cycling with the amplitude
$0.3\,\varepsilon/k_B$.   The nonaffine measures are (a)
$D^2(0,T)>0.04\,\sigma^2$, (b) $D^2(T,T)>0.04\,\sigma^2$, (c)
$D^2(9T,T)>0.04\,\sigma^2$, and (d) $D^2(99T,T)>0.04\,\sigma^2$.}
\label{fig:snapshot_clusters_Tm03_r10m2}
\end{figure}
\begin{figure}[t]
\includegraphics[width=12.cm,angle=0]{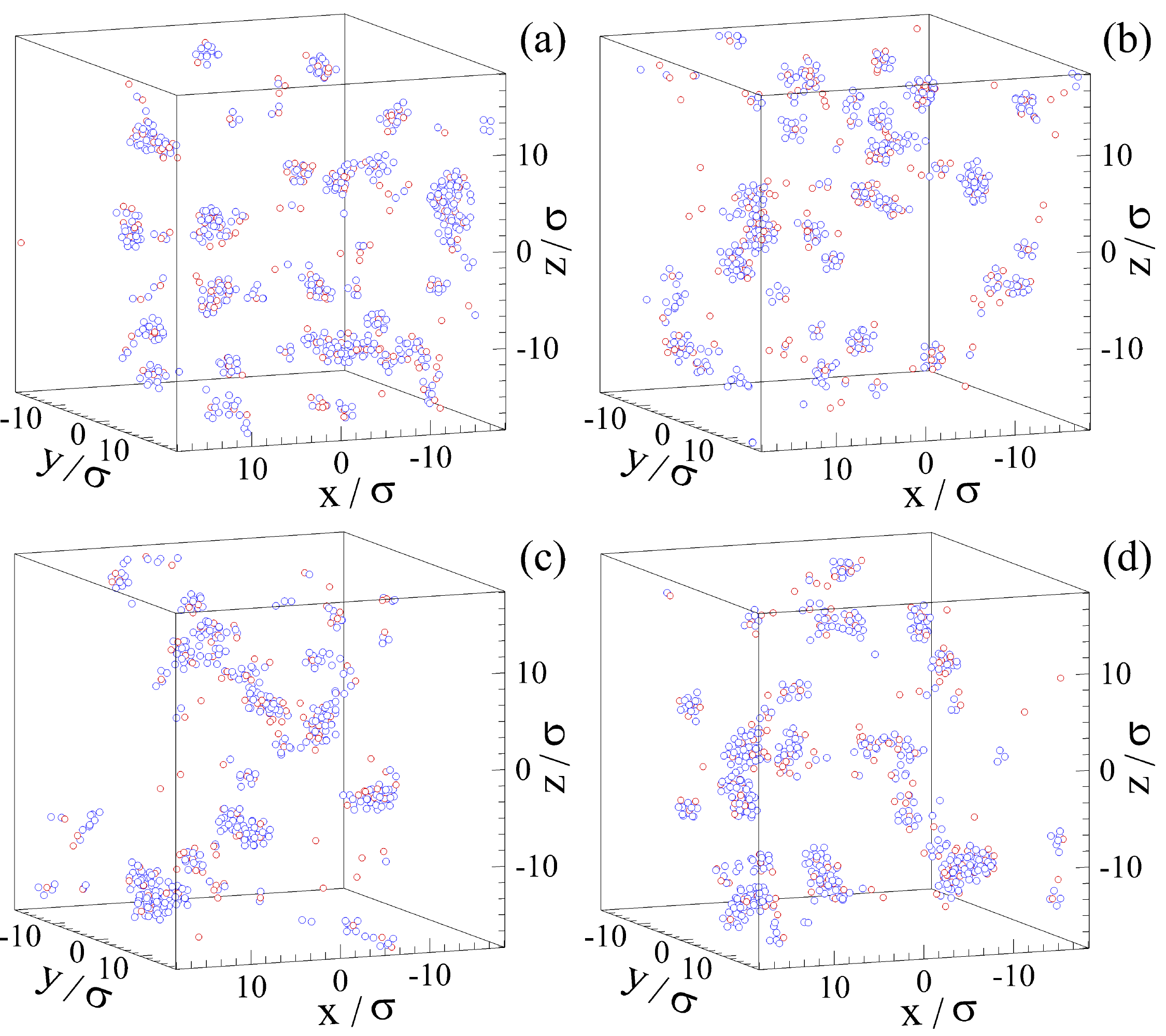}
\caption{(Color online) Positions of atoms with large nonaffile
displacements (a) $D^2(0,T)>0.04\,\sigma^2$, (b)
$D^2(T,T)>0.04\,\sigma^2$, (c) $D^2(9T,T)>0.04\,\sigma^2$, and (d)
$D^2(99T,T)>0.04\,\sigma^2$.  The sample was initially cooled with
the rate of $10^{-5}\varepsilon/k_{B}\tau$ and then thermally cycled
with the amplitude of $0.2\,\varepsilon/k_B$. }
\label{fig:snapshot_clusters_Tm02_r10m5}
\end{figure}
\begin{figure}[t]
\includegraphics[width=12.cm,angle=0]{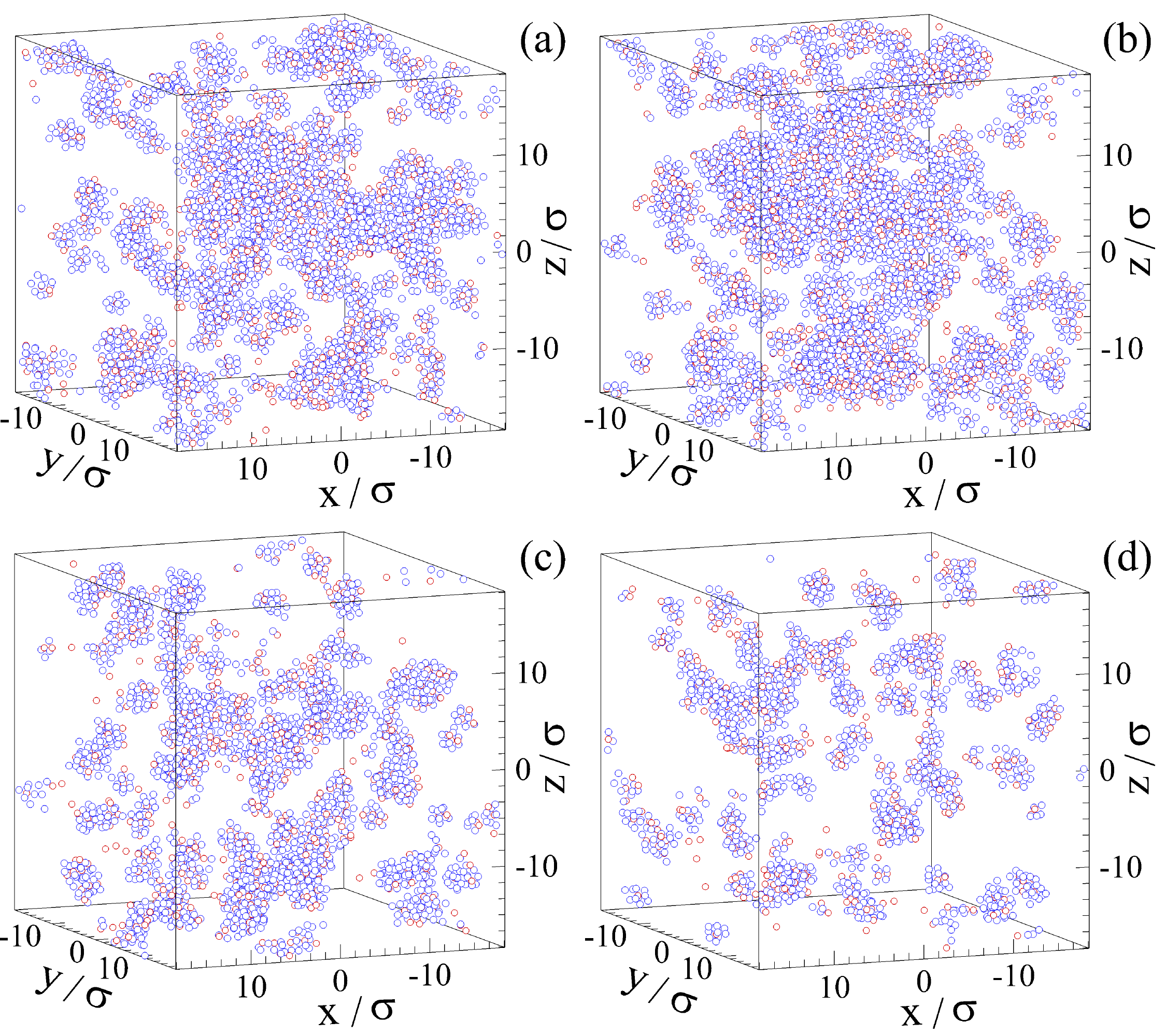}
\caption{(Color online) Snapshots of atoms with relatively large
nonaffine displacements (a) $D^2(0,T)>0.04\,\sigma^2$, (b)
$D^2(T,T)>0.04\,\sigma^2$, (c) $D^2(9T,T)>0.04\,\sigma^2$, and (d)
$D^2(99T,T)>0.04\,\sigma^2$.  The glass is thermally cycled with the
amplitude of $0.3\,\varepsilon/k_B$ and it is initially prepared
with the rate of cooling $10^{-5}\varepsilon/k_{B}\tau$.    }
\label{fig:snapshot_clusters_Tm03_r10m5}
\end{figure}

We next discuss nonaffine displacements in thermally cycled glasses
for two limiting cases of the fastest $10^{-2}\varepsilon/k_{B}\tau$
and slowest $10^{-5}\varepsilon/k_{B}\tau$ cooling rates.  The
representative snapshots of atoms with large nonaffine displacements
during the first, second, 10-th, and 100-th cycles are presented in
Figs.\,\ref{fig:snapshot_clusters_Tm02_r10m2},
\ref{fig:snapshot_clusters_Tm03_r10m2},
\ref{fig:snapshot_clusters_Tm02_r10m5}, and
\ref{fig:snapshot_clusters_Tm03_r10m5} for two cycling amplitudes
$0.2\,\varepsilon/k_B$ and $0.3\,\varepsilon/k_B$. In our analysis,
the nonaffine measure was computed for all atoms during the time
interval $\Delta t = T$, and only atoms with $D^2>0.04\,\sigma^2$
are displayed in
Figs.\,\ref{fig:snapshot_clusters_Tm02_r10m2}--\ref{fig:snapshot_clusters_Tm03_r10m5}.
For comparison, the typical cage size is about $0.1\,\sigma$. It can
be seen in Figs.\,\ref{fig:snapshot_clusters_Tm02_r10m2} and
\ref{fig:snapshot_clusters_Tm03_r10m2} that most of the atoms in the
poorly annealed glass (cooling rate $10^{-2}\varepsilon/k_{B}\tau$)
undergo large nonaffine displacements during the first couple of
cycles, leading to irreversible atom displacements and lower
potential energy states (see Fig.\,\ref{fig:poten_10m2}).  This
process continues for 100 cycles, although the typical size of
clusters of nonaffine displacements is gradually reduced over
consecutive cycles.  Notice that the number of atoms with large
nonaffine displacements after the 100-th cycle is smaller for the
amplitude $0.2\,\varepsilon/k_B$ than for $0.3\,\varepsilon/k_B$,
which suggests that internal stresses due to thermal expansion are
smaller in the former case.

\vskip 0.05in


By contrast, the typical size of clusters of atoms with large
nonaffine displacements for the well annealed glass (cooling rate
$10^{-5}\varepsilon/k_{B}\tau$) depends only weakly on the cycle
number for both temperature amplitudes $0.2\,\varepsilon/k_B$ and
$0.3\,\varepsilon/k_B$, shown in
Figs.\,\ref{fig:snapshot_clusters_Tm02_r10m5} and
\ref{fig:snapshot_clusters_Tm03_r10m5}.   Similar to the poorly
annealed sample, the clusters in the well annealed glass are larger
for the thermal amplitude $0.3\,\varepsilon/k_B$ than for
$0.2\,\varepsilon/k_B$.   Interestingly, despite the appearance of
small-size clusters at the cycling amplitude of
$0.2\,\varepsilon/k_B$ in
Fig.\,\ref{fig:snapshot_clusters_Tm02_r10m5}, the potential energy
minima at the amplitude of $0.2\,\varepsilon/k_B$ remain nearly
equal to the energy level of the glass aged at the constant
temperature $T_{LJ}=0.01\,\varepsilon/k_B$ (see inset to
Fig.\,\ref{fig:poten_10m5}).   We further comment that during
thermal cycling with amplitude $0.4\,\varepsilon/k_B$, most of the
atoms have $D^2>0.04\,\sigma^2$ after one period $T$ for any cooling
rate or cycle number (not shown).   It should be emphasized,
however, that only the well annealed sample, prepared with the
cooling rate of $10^{-5}\varepsilon/k_{B}\tau$, acquires a higher
energy state after 100 cycles with amplitude $0.4\,\varepsilon/k_B$
(see insets to Figs.\,\ref{fig:poten_10m2}--\ref{fig:poten_10m5}).
Finally, the spatial configurations of atoms with large nonaffine
displacements in glasses prepared with intermediate cooling rates,
$10^{-3}\varepsilon/k_{B}\tau$ and $10^{-4}\varepsilon/k_{B}\tau$,
are roughly extrapolated between the two limiting cases, and they
are omitted here for brevity.

\vskip 0.05in

\begin{figure}[t]
\includegraphics[width=12.0cm,angle=0]{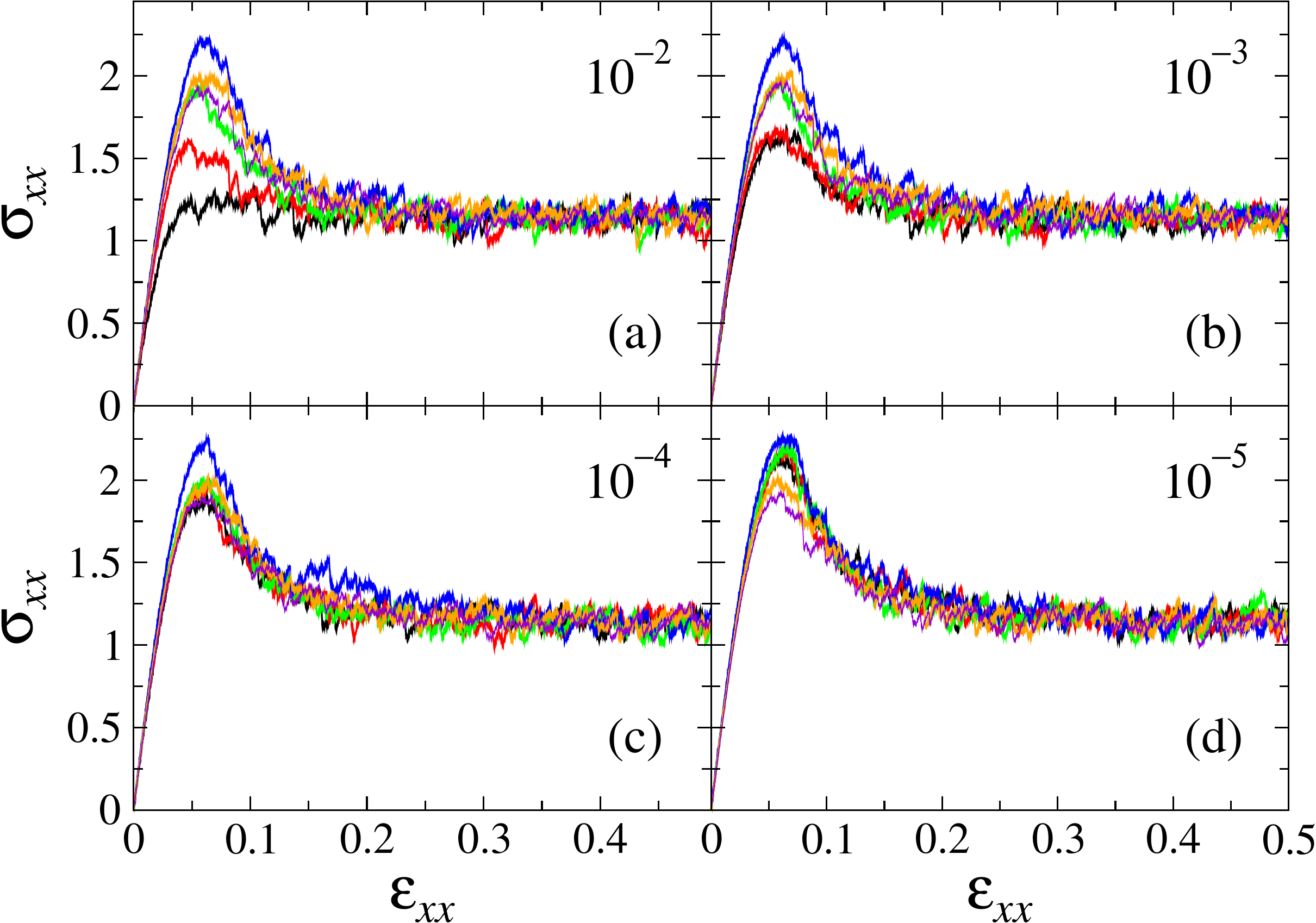}
\caption{(Color online) The dependence of tensile stress
$\sigma_{xx}$ (in units of $\varepsilon\sigma^{-3}$) as a function
of strain for samples obtained with cooling rates (a)
$10^{-2}\varepsilon/k_{B}\tau$, (b) $10^{-3}\varepsilon/k_{B}\tau$,
(c) $10^{-4}\varepsilon/k_{B}\tau$, and (d)
$10^{-5}\varepsilon/k_{B}\tau$. The stress response for glasses aged
at the constant temperature of $0.01\,\varepsilon/k_B$ is denoted by
the black curves. The stress-strain curves are computed after 100
periods of thermal cycling with the amplitudes
$0.1\,\varepsilon/k_B$ (red), $0.2\,\varepsilon/k_B$ (green),
$0.3\,\varepsilon/k_B$ (blue), $0.4\,\varepsilon/k_B$ (orange), and
$0.435\,\varepsilon/k_B$ (violet). The strain rate is
$\dot{\varepsilon}_{xx}=10^{-5}\,\tau^{-1}$.}
\label{fig:stress_strain}
\end{figure}

We find that the thermal treatment of binary glasses can
significantly affect their mechanical properties.  The stress-strain
curves for different cycling amplitudes are presented in
Fig.\,\ref{fig:stress_strain} for cooling rates
$10^{-2}\varepsilon/k_{B}\tau$, $10^{-3}\varepsilon/k_{B}\tau$,
$10^{-4}\varepsilon/k_{B}\tau$, and $10^{-5}\varepsilon/k_{B}\tau$.
All samples were strained with the rate
$\dot{\varepsilon}_{xx}=10^{-5}\,\tau^{-1}$ after 100 thermal
cycles. For reference, the data for glasses aged at the constant
temperature $T_{LJ}=0.01\,\varepsilon/k_B$ during the time interval
$10^6\tau=100\,T$ are also reported in
Fig.\,\ref{fig:stress_strain}.   As expected, upon decreasing
cooling rate, a pronounced peak at the yield strain is developed for
glasses aged at $T_{LJ}=0.01\,\varepsilon/k_B$.   Furthermore, it
can be seen in Fig.\,\ref{fig:stress_strain} that the dependence of
the yield stress on the cycling amplitude is nonmonotonic, and the
highest peak appears at $T_{LJ}=0.3\,\varepsilon/k_B$ for all
cooling rates. This behavior correlates well with the occurrence of
the lowest minima of the potential energy at the amplitude of
$0.3\,\varepsilon/k_B$ shown in the insets to
Figs.\,\ref{fig:poten_10m2}--\ref{fig:poten_10m5}.

\vskip 0.05in

\begin{figure}[t]
\includegraphics[width=12.cm,angle=0]{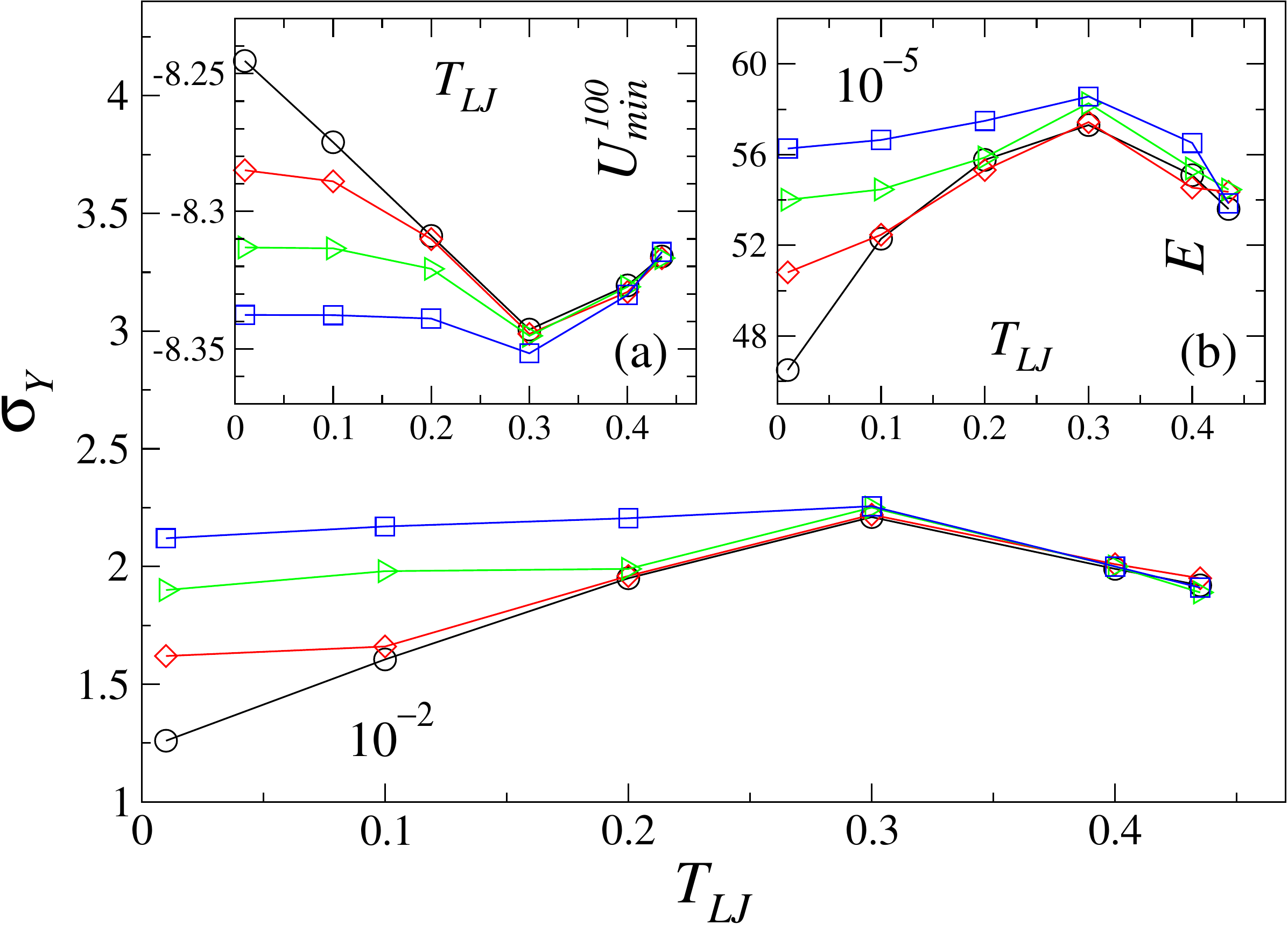}
\caption{(Color online) The peak value of the stress overshoot
$\sigma_Y$ (in units of $\varepsilon\sigma^{-3}$) versus the
amplitude of thermal cycling for cooling rates
$10^{-2}\varepsilon/k_{B}\tau$ (black curve),
$10^{-3}\varepsilon/k_{B}\tau$ (red curve),
$10^{-4}\varepsilon/k_{B}\tau$ (green curve), and
$10^{-5}\varepsilon/k_{B}\tau$ (blue curve).  The minimum of the
potential energy after 100 cycles is shown in the inset (a) for
different cooling rates (the colorcode is the same). The inset (b)
shows the elastic modulus $E$ (in units of $\varepsilon\sigma^{-3}$)
as a function of the temperature amplitude for the same samples. }
\label{fig:yield_stress_E}
\end{figure}

The summary of the data for the yielding peak, $\sigma_Y$, the
elastic modulus, $E$, and the potential energy minimum, $U(100\,T)$,
as a function of the cycling amplitude is presented in
Fig.\,\ref{fig:yield_stress_E}. As is evident, both $\sigma_Y$ and
$E$ increase, while $U(100\,T)$ is reduced, when $T_{LJ}$ approaches
$0.3\,\varepsilon/k_B$ from below.  It can be observed that at
temperature amplitudes above $0.3\,\varepsilon/k_B$, the values of
the yielding peak and the potential energy minimum after 100 cycles
are only weakly dependent on the cooling rate.  Note also that the
yielding peak for the rejuvenated glass (cooling rate
$10^{-5}\varepsilon/k_{B}\tau$) at the cycling amplitude
$0.4\,\varepsilon/k_B$ is smaller than $\sigma_Y$ for the same
sample aged at $T_{LJ}=0.01\,\varepsilon/k_B$. This behavior
correlates well with the dependence of $U(100\,T)$ for the well
annealed sample, shown in the inset (a) to
Fig.\,\ref{fig:yield_stress_E}. Furthermore, the numerical results
in Fig.\,\ref{fig:yield_stress_E} suggest that the peak value of the
stress overshoot for poorly annealed glasses that are cycled with
amplitudes below $0.3\,\varepsilon/k_B$ represents a lower bound for
$\sigma_Y$ measured in more slowly annealed glasses. Plotted in the
inset (b) to Fig.\,\ref{fig:yield_stress_E}, the elastic modulus
shows a similar dependence on the thermal amplitude and cooling
rate, although the data appear to be more scattered at large
amplitudes. Altogether, these results demonstrate that the elastic
modulus and the stress overshoot increase in binary glasses
subjected to multiple periods of heating and cooling up to a certain
temperature. The effect is most pronounced at the maximum
temperature $T_{LJ}\approx0.69\,T_c$.

\section{Conclusions}

In summary, molecular dynamics simulations were carried out to study
structural relaxation and mechanical properties of a model glass
after thermal cycling with different amplitudes. We considered a
binary Lennard-Jones mixture, which was initially annealed from the
liquid state to a temperature well below the glass transition. With
increasing cooling rate, the glasses settle into more shallow
potential energy minima and become less dense. After annealing, the
thermal treatment was imposed during one hundred cycles of heating
and cooling at constant pressure.  We found that depending on the
cooling rate and cycling amplitude, the glasses can be either
overaged or rejuvenated by periodic temperature variations. Thus,
the potential energy in quickly annealed glasses decreases during
thermal cycling, while the potential energy in slowly annealed
glasses increases at large cycling amplitudes.  It was further shown
that the lowest energy minima are attained at a particular
temperature amplitude for quickly and slowly annealed glasses. In
all cases, the structural changes leading to different energy states
occur via collective nonaffine displacements of atoms, and the
typical size of clusters of mobile atoms is reduced with decreasing
cooling rate or temperature amplitude and with increasing cycle
number. In turn, the mechanical properties were probed via uniaxial
tensile loading at constant pressure after the thermal treatment.
The simulation results demonstrated that the elastic modulus and the
peak value of the stress overshoot become larger with increasing
cycling amplitude up to about two-thirds of the glass transition
temperature. In this range, the variation of the yield peak as a
function of the temperature amplitude for quickly annealed glasses
provides a lower bound for the maximum stress in more slowly
annealed glasses.

\section*{Acknowledgments}

Financial support from the National Science Foundation (CNS-1531923)
is gratefully acknowledged. The article was prepared within the
framework of the Basic Research Program at the National Research
University Higher School of Economics (HSE) and supported within the
framework of a subsidy by the Russian Academic Excellence Project
`5-100'. The molecular dynamics simulations were performed using the
LAMMPS numerical code developed at Sandia National
Laboratories~\cite{Lammps}. Computational work in support of this
research was performed at Wright State University's Computing
Facility and the Ohio Supercomputer Center.


%

%

%

%
%

%
%

%
%

%
%

%

%

%

%

%

%

\bibliographystyle{prsty}

\begin{thebibliography}{99}


\bibitem{Kruzic16}     J.~J. Kruzic,
                       Bulk Metallic Glasses as Structural Materials: A Review,
                       Adv. Eng. Mater. {\bf 18}, 1308 (2016).


\bibitem{Argon79}      A.~S. Argon,
                       Plastic deformation in metallic glasses,
                       Acta Metall. {\bf 27}, 47 (1979).

\bibitem{Spaepen77}    F. Spaepen,
                       A microscopic mechanism for steady state inhomogeneous flow in metallic glasses,
                       Acta Metall. {\bf 25}, 407 (1977).

\bibitem{Ketov15}      S.~V. Ketov, Y.~H. Sun, S. Nachum, Z. Lu, A. Checchi, A.~R. Beraldin, H.~Y. Bai,
                       W.~H. Wang, D.~V. Louzguine-Luzgin, M.~A. Carpenter, and A.~L. Greer,
                       Rejuvenation of metallic glasses by non-affine thermal strain,
                       Nature {\bf 524}, 200 (2015).

\bibitem{Lu18}         W. Song, X. Meng, Y. Wu, D. Cao, H. Wang, X. Liu, X. Wang, and Z. Lu,
                       Improving plasticity of the Zr$_{46}$Cu$_{46}$Al$_8$ bulk metallic glass
                       via thermal rejuvenation,
                       Science Bulletin {\bf 63}, 840 (2018).

\bibitem{Kerscher18}   D. Grell, F. Dabrock, and E. Kerscher,
                       Cyclic cryogenic pretreatments influencing the mechanical
                       properties of a bulk glassy Zr-based alloy,
                       Fatigue Fract. Eng. Mater. Struct. {\bf 41}, 1330 (2018).

\bibitem{Saida18}      W. Guoa, R. Yamada, and J. Saida,
                       Rejuvenation and plasticization of metallic glass by deep cryogenic cycling treatment,
                       Intermetallics {\bf 93}, 141  (2018).


\bibitem{Ketov18}      S.~V. Ketov, A.~S. Trifonov, Y.~P. Ivanov, A.~Yu. Churyumov, A.~V. Lubenchenko,
                       A.~A. Batrakov, J. Jiang, D.~V. Louzguine-Luzgin, J. Eckert, J. Orava, and A.~L. Greer,
                       On cryothermal cycling as a method for inducing structural changes in metallic glasses,
                       NPG Asia Materials {\bf 10}, 137 (2018).

\bibitem{Hufnagel15}   T.~C. Hufnagel,
                       Metallic glasses: Cryogenic rejuvenation,
                       Nature Materials {\bf 14}, 867 (2015).


\bibitem{Barrat18}     B. Shang, P. Guan, and J.-L. Barrat,
                       Role of thermal expansion heterogeneity in the cryogenic rejuvenation of metallic glasses,
                       arXiv:1807.00760 (2018).


\bibitem{Lacks04}      D.~J. Lacks and M.~J. Osborne,
                       Energy landscape picture of overaging and rejuvenation in a sheared glass,
                       Phys. Rev. Lett. {\bf 93}, 255501 (2004).




\bibitem{Priezjev13}   N.~V. Priezjev,
                       Heterogeneous relaxation dynamics in amorphous materials under cyclic loading,
                       Phys. Rev. E {\bf 87}, 052302 (2013).

\bibitem{Sastry13}     D. Fiocco, G. Foffi, and S. Sastry,
                       Oscillatory athermal quasistatic deformation of a model glass,
                       Phys. Rev. E {\bf 88}, 020301(R) (2013).

\bibitem{Reichhardt13} I. Regev, T. Lookman, and C. Reichhardt,
                       Onset of irreversibility and chaos in amorphous solids under periodic shear,
                       Phys. Rev. E {\bf 88}, 062401 (2013).

\bibitem{Priezjev14}   N.~V. Priezjev,
                       Dynamical heterogeneity in periodically deformed polymer glasses,
                       Phys. Rev. E {\bf 89}, 012601 (2014).

\bibitem{IdoNature15}  I. Regev, J. Weber, C. Reichhardt, K.~A. Dahmen, and T. Lookman,
                       Reversibility and criticality in amorphous solids,
                       Nat. Commun. {\bf 6}, 8805 (2015).

\bibitem{Priezjev16}   N.~V. Priezjev,
                       Reversible plastic events during oscillatory deformation of amorphous solids,
                       Phys. Rev. E {\bf 93}, 013001 (2016).


\bibitem{Kawasaki16}   T. Kawasaki and L. Berthier,
                       Macroscopic yielding in jammed solids is accompanied by a non-equilibrium
                       first-order transition in particle trajectories,
                       Phys. Rev. E {\bf 94}, 022615 (2016).

\bibitem{Priezjev16a}  N.~V. Priezjev,
                       Nonaffine rearrangements of atoms in deformed and quiescent binary glasses,
                       Phys. Rev. E {\bf 94}, 023004 (2016).


\bibitem{Sastry17}     P. Leishangthem, A.~D.~S. Parmar, and S. Sastry,
                       The yielding transition in amorphous solids under oscillatory shear deformation,
                       Nat. Commun. {\bf 8}, 14653 (2017).


\bibitem{Priezjev17}   N.~V. Priezjev,
                       Collective nonaffine displacements in amorphous materials during large-amplitude oscillatory shear,
                       Phys. Rev. E {\bf 95}, 023002 (2017).


\bibitem{OHern17}      M. Fan, M. Wang, K. Zhang, Y. Liu, J. Schroers, M.~D. Shattuck, and C.~S. O'Hern,
                       The effects of cooling rate on particle rearrangement statistics:
                       Rapidly cooled glasses are more ductile and less reversible,
                       Phys. Rev. E {\bf 95}, 022611 (2017).

\bibitem{Hecke17}      S. Dagois-Bohy, E. Somfai, B.~P. Tighe, and M. van Hecke,
                        Softening and yielding of soft glassy materials,
                        Soft Matter {\bf 13}, 9036 (2017).

\bibitem{Priezjev18}   N.~V. Priezjev,
                       Molecular dynamics simulations of the mechanical annealing process in
                       metallic glasses: Effects of strain amplitude and temperature,
                       J. Non-Cryst. Solids {\bf 479}, 42 (2018).

\bibitem{Priezjev18a}  N.~V. Priezjev,
                       The yielding transition in periodically sheared binary glasses at finite temperature,
                       Comput. Mater. Sci. {\bf 150}, 162 (2018).


\bibitem{Sastry18}     P. Das, A.~D.~S. Parmar, and S. Sastry,
                       Annealing glasses by cyclic shear deformation,
                       arXiv:1805.12476 (2018).





\bibitem{KobAnd95}     W. Kob and H.~C. Andersen,
                       Testing mode-coupling theory for a supercooled binary Lennard-Jones mixture:
                       The van Hove correlation function,
                       Phys. Rev. E {\bf 51}, 4626 (1995).

\bibitem{Weber85}      T.~A. Weber and F.~H. Stillinger,
                       Local order and structural transitions in amorphous metal-metalloid alloys,
                       Phys. Rev. B {\bf 31}, 1954 (1985).


\bibitem{Allen87}      M.~P. Allen and D.~J. Tildesley,
                       {\it Computer Simulation of Liquids} (Clarendon, Oxford, 1987).


\bibitem{Lammps}       S.~J. Plimpton,
                       Fast parallel algorithms for short-range molecular dynamics,
                       J. Comp. Phys. {\bf 117}, 1 (1995).

\bibitem{KobBar97}     W. Kob and J.-L. Barrat,
                       Aging Effects in a Lennard-Jones Glass,
                       Phys. Rev. Lett. {\bf 78}, 4581 (1997).

\bibitem{KobBar00}     W. Kob and J.-L. Barrat,
                       Fluctuations, response and aging dynamics in a simple
                       glass-forming liquid out of equilibrium,
                       Eur. Phys. J. B {\bf 13}, 319 (2000).

\bibitem{Falk98}       M.~L. Falk and J.~S. Langer,
                       Dynamics of viscoplastic deformation in amorphous solids,
                       Phys. Rev. E {\bf 57}, 7192 (1998).

\bibitem{NVP18strload} N.~V. Priezjev,
                       Slow relaxation dynamics in binary glasses during stress-controlled,
                       tension-compression cyclic loading,
                       Comput. Mater. Sci. {\bf 153}, 235 (2018).






\end{thebibliography}

\end{document}